\documentclass[twocolumn,prl,showpacs]{revtex4}

\usepackage{graphicx}
\usepackage{dcolumn}
\usepackage{bm}
\usepackage{amsmath}

\begin{document}

\title{Anisotropic superconducting properties of aligned Sm$_{0.95}$La$_{0.05}$FeAsO$_{0.85}$F$_{0.15}$ microcrystalline
powder}
\author{B. C. Chang$^{1}$, C. H. Hsu$^{1}$, Y. Y. Hsu$^{2}$, Z. Wei$^{3}$,  K. Q. Ruan$^{3}$, X. G. Li$^{3}$, and H. C. Ku$^{1}$}
\email{hcku@phys.nthu.edu.tw} \affiliation{$^{1}$Department of
Physics, National Tsing Hua University, Hsinchu 30013, Taiwan
\\$^{2}$Department of Physics, National Taiwan Normal
University, Taipei 10610, Taiwan
\\$^{3}$Hefei National Laboratory for Physical Sciences at Microscale and Department of Physics, University of Science and Technology of China, Hefei 230026, China}
\date{\today}

\pacs{74.72.-h, 74.25.Ha}

\begin{abstract}
The Sm$_{0.95}$La$_{0.05}$FeAsO$_{0.85}$F$_{0.15}$ compound is a
quasi-2D layered superconductor with a superconducting transition
temperature T$_c$ = 52 K. Due to the Fe spin-orbital related
anisotropic exchange coupling (antiferromagnetic or ferromagnetic
fluctuation), the tetragonal microcrystalline powder can be aligned
at room temperature using the field-rotation method where the
tetragonal $\it{ab}$-plane is parallel to the aligned magnetic field
B$_{a}$ and $\it{c}$-axis along the rotation axis. Anisotropic
superconducting properties with anisotropic diamagnetic ratio
$\chi_{c}$/$\chi_{ab}\sim$ 2.4 + 0.6 was observed from low field
susceptibility $\chi$(T) and magnetization M(B$_{a}$). The
anisotropic low-field phase diagram with the variation of lower
critical field gives a zero-temperature penetration depth
$\lambda_{c}$(0) = 280 nm and $\lambda_{ab}$(0) = 120 nm. The
magnetic fluctuation used for powder alignment at 300 K may be
related with the pairing mechanism of superconductivity at lower
temperature.

\end{abstract}

\maketitle

High-T$_{c}$ superconductivity with transition temperature T$_{c}$
up to 55 K were reported in the newly discovered iron-based
RFeAsO$_{1-x}$F$_{x}$ (rare earth R = La, Ce, Pr, Nd or Sm) system
\cite{p1,p2,p3,p4,p5,p6,p7,p8,p9,p10,p11,p12}. The ZrCuAsSi-type
tetragonal structure (space group P4/nmm) is a layered structure
where the metallic FeAs layer is separated by the insulating
RO$_{1-x}$F$_{x}$ layer. The discovery of the iron-based
superconductor has generated enormous interest since these compounds
are the first non-cuprate high-T$_{c}$ superconductors with T$_{c}$
higher than 50 K. The parent compound LaFeAsO is a normal semi-metal
which shows a Fermi surface nesting or spin Peierls instability
below 150 K with a tetragonal to orthorhombic structural transition,
accompanied by a spin density wave (SDW) type antiferromagnetic
order \cite{p6}. Electron doping to the FeAs layer through
F$^{-}$-substitution in the O$^{2-}$ site or O$^{2-}$-deficiency
suppresses both the magnetic order and the structural distortion in
favor of superconductivity \cite{p1,p3}. On the other hand, hole
doping through Sr$^{2+}$-substitution in the R$^{3+}$ site gives
similar effect \cite{p4}. Therefore, like the high-T$_{c}$ cuprate
systems, the superconductivity in these iron-based compounds occurs
in close proximity to a long range antiferromagnetic ground state.

Since the FeAs layer is believed to be the superconducting layer of
the RFeAsO$_{1-x}$F$_{x}$ system, studies on the anisotropic
properties are crucial for understanding this new iron-based system.
High-quality single crystal is essential for detailed in-depth
studies \cite{p9,p12}. However, the anisotropic superconducting
properties can be easily obtained using a much simpler way. In this
report, we use the field-rotation alignment method to align the
T$_{c}$ = 52 K Sm$_{0.95}$La$_{0.05}$FeAsO$_{0.85}$F$_{0.15}$
microcrystalline powder at room temperature.
\begin{figure}
\includegraphics[scale=0.3]{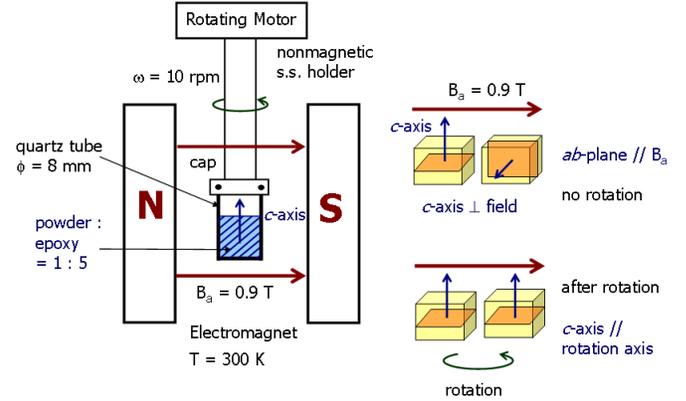}
\caption{\label{label}  Block diagram for field-rotation powder
alignment method with $\it{ab}$-plane along the aligned magnetic
field B$_{a}$ and $\it{c}$-axis along the rotation axis.}
\end{figure}
The polycrystalline Sm$_{0.95}$La$_{0.05}$FeAsO$_{0.85}$F$_{0.15}$
was prepared by conventional solid state reaction. First the SmAs
powder was prepared by reacting Sm and As powders at 650$^{\circ}$C
for about 5 hours, and fine powders of SmAs, FeAs, Fe, LaF$_3$,
Fe$_2$O$_3$ were mixed together according to the stoichiometric
ratio of Sm$_{0.95}$La$_{0.05}$FeAsO$_{0.85}$F$_{0.15}$, ground
thoroughly, and then pressed into pellets. The pellets were wrapped
in Ta foil to be sealed in an evacuated quartz tube, and annealed at
1160$^{\circ}$C for about 50 hours. The
Sm$_{0.95}$La$_{0.05}$FeAsO$_{0.85}$F$_{0.15}$ powders with an
average microcrystalline grain size of 1-10 $\mu$m were mixed with
epoxy (4-hour curing time) in a quartz tube ($\phi$ = 8 mm) with a
powder:epoxy ratio of 1:5. The quartz tube was immediately placed in
a 0.9-T electromagnet and rotated at a speed of 10 rpm with the
rotation axis perpendicular to the aligned magnetic field B$_{a}$
(Fig. 1). Since the tetragonal $\it{ab}$-plane of
Sm$_{0.95}$La$_{0.05}$FeAsO$_{0.85}$F$_{0.15}$ microcrystalline is
aligned along B$_{a}$ from X-ray diffraction study, the rotation
perpendicular to B$_{a}$ will force the microcrystalline
$\it{c}$-axis to align along the rotation axis.

\begin{figure}
\includegraphics{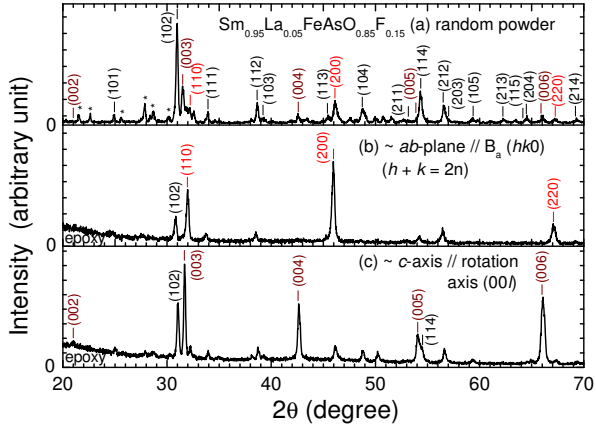}
\caption{\label{label} Powder X-ray diffraction patterns for
Sm$_{0.95}$La$_{0.05}$FeAsO$_{0.85}$F$_{0.15}$. (a) random powder,
(b) $\it{ab}$-plane aligned along B$_{a}$, (c) $\it{c}$-axis aligned
along rotation axis ($\perp$ B$_{a}$).}
\end{figure}
The X-ray diffraction patterns for
Sm$_{0.95}$La$_{0.05}$FeAsO$_{0.85}$F$_{0.15}$ random powder,
partially $\it{ab}$-plane aligned along B$_{a}$, and partially
$\it{c}$-axis aligned along the rotation axis are shown collectively
in Fig. 2. The unaligned random powder diffraction pattern gives a
slightly larger tetragonal lattice parameters of $\it{a}$ =
0.3936(3) nm and $\it{c}$ = 0.8495(8) nm with 5$\%$ La doping than
SmFeAsO$_{0.85}$F$_{0.15}$ \cite{p5}. At least 10$\%$ impurities
were observed as marked by asterisk in the diffraction pattern. For
$\it{ab}$-plane aligned along B$_{a}$, enhanced ($\it{hk}$0)
diffraction lines ($\it{h}$ + $\it{k}$ = 2n) were observed. The lack
of forbidden line (210) is consistent with the space group P4/nmm.
For $\it{c}$-axis aligned along the rotation axis, enhanced
(00$\it{l}$) diffraction lines were observed. The imperfect
alignment (80-90 $\%$) with the appearance of major (102) line may
be due to intrinsic weak magnetic anisotropy, low aligned field (0.9
T), imperfect powder preparation or alignment procedure. The
diffraction pattern for nonsuperconducting
SmFeAsO$_{0.95}$F$_{0.05}$ aligned powder gives similar result with
$\it{ab}$-plane aligned along B$_{a}$.

The field-rotation alignment method at room temperature is based on
magnetic anisotropy due to the spin-orbital related anisotropic
exchange coupling. There are two major contributions of magnetic
moment at room temperature $\chi \sim \chi$(Fe) + $\chi$(Sm). The
large Sm$^{3+}$ (4f$^{5}$5s$^{2}$5p$^{6}$, S = 5/2, L = 5, J = L - S
= 5/2) localized moment in the insulating
(Sm$_{0.95}$La$_{0.05}$)(O$_{0.85}$F$_{0.15}$) layer has weak
anisotropic exchange coupling. On the other hand, the Fe$^{2+}$
(3d$^{6}$) itinerant moment in the metallic FeAs layer is rather
small from powder neutron diffraction data and band structure
calculation. The Fe$^{2+}$ feels the distorted FeAs$_{4}$
tetrahedral crystal field with three low-lying manifold and two
up-lying manifold (d$_{xz}$ and d$_{yz}$). The six 3$\it{d}$
electrons are distributed in three quasi-2D hole-like bands and two
electron-like bands with Fermi surfaces from five 3$\it{d}$
orbitals. Anisotropic exchange coupling (magnetic fluctuation)
occurs from strong Fe-As 3d$_{xz,yz}$-4p hybridization and Fe-Fe
3d$_{xy}$-3d$_{xy}$ direct electron hopping between neighboring Fe
atoms in the tetragonal basal $\it{ab}$-plane \cite{p6,p10,p11}. In
the aligned magnetic field, anisotropic Fe orbitals are tied to the
spin direction, and a strong spin-orbital related anisotropic
exchange coupling at 300 K should dominate the magnetic alignment.
Due to the quasi-2D FeAs layer structure, anisotropic Fe magnetic
susceptibility $\chi_{ab}$(Fe) $>$ $\chi_{c}$(Fe) is expected to be
the dominant factor for $\it{ab}$-plane alignment along B$_{a}$ at
300 K. Indeed, anisotropic room temperature magnetization for
Sm$_{0.95}$La$_{0.05}$FeAsO$_{0.85}$F$_{0.15}$ aligned powder was
observed with weak magnetic anisotropy ratio $\chi_{ab}/\chi_{c}
\sim$ 1.2 up to 7 T. In the low field region, a cross-over from
$\chi_{c} \ge \chi_{ab}$ to $\chi_{c} \le \chi_{ab}$ at high field
with nonlinear behavior may be caused by the magnetic impurities.

\begin{figure}
\includegraphics{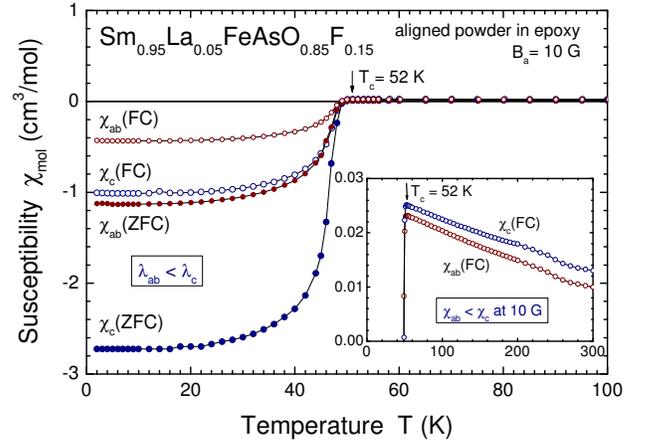}
\caption{\label{label} Anisotropic low-field FC and ZFC
susceptibility $\chi_{ab}$(T) and $\chi_{c}$(T) for aligned
Sm$_{0.95}$La$_{0.05}$FeAsO$_{0.85}$F$_{0.15}$.}
\end{figure}
Magnetization and magnetic susceptibility data were collected with a
Quantum Design 1-T $\mu$-metal shielded MPMS$_2$ or a 7-T MPMS
superconducting quantum interference device (SQUID) magnetometer
from 2 K to 300 K. The anisotropic temperature dependence of molar
magnetic susceptibility $\chi_{ab}$(T) and $\chi_{c}$(T) for aligned
powder Sm$_{0.95}$La$_{0.05}$FeAsO$_{0.85}$F$_{0.15}$ with applied
field along the $\it{ab}$-plane and $\it{c}$-axis are shown
collectively in Fig. 3. For aligned dispersed microcrystalline in
low applied field of 10 G, both zero-field-cooled (ZFC) and
field-cooled (FC) data revealed a sharp superconducting transition
temperature T$_{c}$ of 52 K, which is identical to the measured
T$_{c}$ from  bulk polycrystalline sample. Large ZFC intragrain
Meissner shielding signals were observed with an almost constant
$\chi_{c}$ = -2.72 cm$^{3}$/mol and $\chi_{ab}$ = -1.13 cm$^{3}$/mol
up to 20 K. The anisotropic diamagnetic parameter $\gamma$ =
$\chi_{c}$/$\chi_{ab}$ of 2.4 was deduced for aligned
microcrystalline. The FC flux-trapped signal gave the same
anisotropic parameter $\gamma$ of 2.4. Considering the imperfect
alignment factor (80-90$\%$), a larger anisotropic diamagnetic
parameter $\gamma$ = 2.4 + 0.6 was derived. For temperature above
T$_{c}$, normal state paramagnetic anisotropy ratio
$\chi_{c}$/$\chi_{ab}$ of 1.3 was observed in 10 G at 300 K, which
is consistent with low field magnetization at 300 K and may be
caused by magnetic impurities. No trace of Curie-Weiss-like behavior
was observed in the normal state.

\begin{figure}
\includegraphics{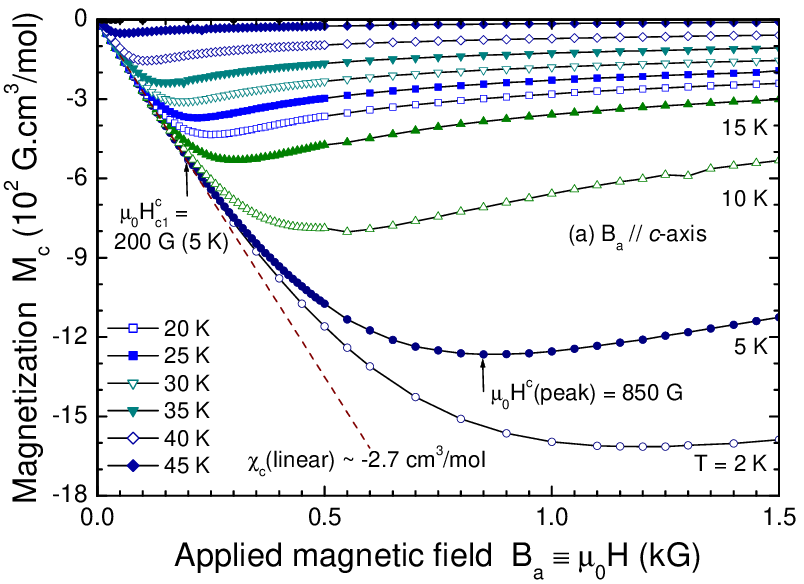}
\includegraphics{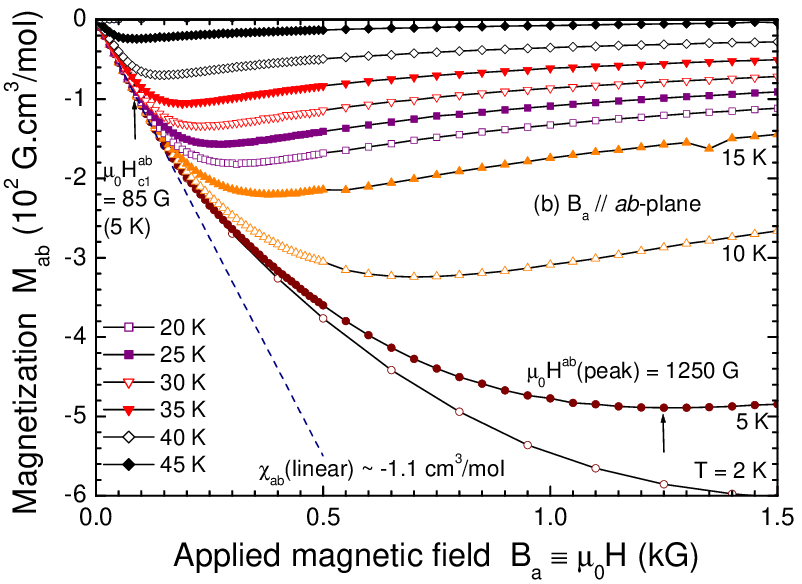}
\caption{\label{label} Temperature variation of (a) $\it{c}$-axis
superconducting initial magnetization M$_{c}$(B$_{a}$) and (b)
$\it{ab}$-plane superconducting initial magnetization
M$_{ab}$(B$_{a}$) for aligned
Sm$_{0.95}$La$_{0.05}$FeAsO$_{0.85}$F$_{0.15}$ powder.}
\end{figure}
The temperature variation of superconducting initial magnetization
curves with applied field B$_{a}$ along $\it{c}$-axis,
M$_{c}$(B$_{a}$,T), and $\it{ab}$-plane, M$_{ab}$(B$_{a}$,T), for
aligned Sm$_{0.95}$La$_{0.05}$FeAsO$_{0.85}$F$_{0.15}$ powder from 2
K to 45 K are shown in Fig. 4(a) and 4(b), respectively. For
example, at T = 5 K, the maximum diamagnetic signal was observed at
magnetization peak field $\mu_{0}$H$^{c}$(peak) of 850 G for M$_{c}$
and a larger $\mu_{0}$H$^{ab}$(peak) of 1250 G for M$_{ab}$,
indicating anisotropic pinning for different applied field
directions. The peak field corresponds to the field that magnetic
flux penetrates into the center of aligned superconducting
microcrystallines forming a vortex glass where vortex are mainly
pinned by impurities and imperfections. At the low-field region,
linear variation of magnetization as an evidence of Meissner state
was clearly observed for both directions \cite{p13}. The low-field
linear slopes of initial magnetization are consistent with the
low-field (B$_{a}$ = 10 G) susceptibility as measured in Fig. 3.
Since the lower critical field $\mu_{0}$H$_{c1}$ is the field flux
start to penetrate into the superconductor forming mixed state , the
magnetization M(B$_{a}$) deviates from the linear response of
Meissner state. Using 10-G/5-K linear magnetic susceptibility
$\chi_{c}$ = -2.7 cm$^{3}$/mol and $\chi_{ab}$ = -1.1 cm$^{3}$/mol
in Fig. 3, lower critical field $\mu_{0}$H$_{c1}^{c}$ of 200 G and
$\mu_{0}$H$_{c1}^{ab}$ of 85 G were deduced from the linear
susceptibility extrapolation lines.

The low temperature, low magnetic field anisotropic phase diagram
for aligned Sm$_{0.95}$La$_{0.05}$FeAsO$_{0.85}$F$_{0.15}$ powder is
shown in Fig. 5(a). The zero temperature lower critical field
$\mu_{0}$H$_{c1}^{c}$(0) of 230 G and $\mu_{0}$H$_{c1}^{ab}$(0) of
95 G were extrapolated. The temperature dependence of anisotropic
penetration depth $\lambda_{ab}$(T) and $\lambda_{c}$(T) are shown
in Fig. 5(b), using the anisotropic formula $\mu_{0}$H$_{c1}^{c}$ =
$\Phi_{0}$/2$\pi\lambda_{ab}^{2}$ and $\mu_{0}$H$_{c1}^{ab}$ =
$\Phi_{0}$/2$\pi\lambda_{ab}\lambda_{c}$, where $\Phi_{0}$ is the
flux quantum. The zero temperature penetration depths
$\lambda_{c}$(0) = 280 nm and $\lambda_{ab}$(0) = 120 nm were
extrapolated. These values reflect the quasi-2D FeAs layer structure
and are smaller than the microcrystalline grain size of $\sim$ 1-10
$\mu$m. The anisotropy parameter $\gamma$ derived from the
penetration depth ratio $\lambda_{c}(0)/\lambda_{ab}(0)$ of 2.33 is
smaller than the reported value from single crystal by penetration
depth ($\sim$ 4) \cite{p12} or $\mu_{0}$H$_{c2}$ ($\sim$ 4.34-4.9)
\cite{p9}. However, the differences of anisotropy parameter $\gamma$
may be attributed to the different demagnetization field in
slab-shaped single crystal with different applied field directions
when the geometric shape is considered, i.e. the slab shape of
single crystals and the sphere-like shape of the aligned
microcrystalline in this work.

\begin{figure}
\includegraphics{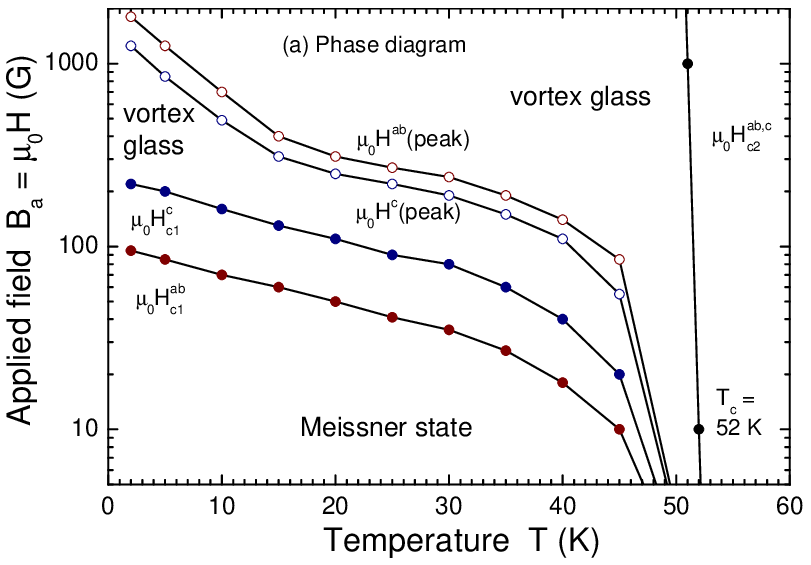}
\includegraphics{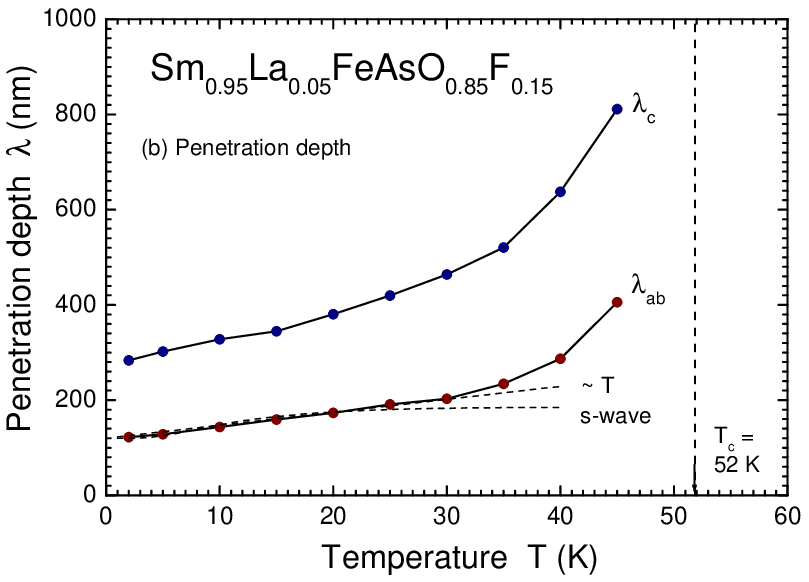}
\caption{\label{label}  (a) Low field anisotropic phase diagram
$\mu_{0}$H(T) for aligned
Sm$_{0.95}$La$_{0.05}$FeAsO$_{0.85}$F$_{0.15}$ powder. (b)
Temperature dependence of anisotropic penetration depth
$\lambda_{ab}$(T) and $\lambda_{c}$(T).}
\end{figure}
Since the band structure calculations suggested quasi-2D nature of
the Fermi surface \cite{p10,p11} in the layered Fe-based compounds,
the $\it{ab}$-plane penetration depth behavior was expected to
provide more intrinsic information than $\it{c}$-axis. The low
temperature (2-25 K) penetration depth $\lambda_{ab}$(T) can be
roughly fitted by a BCS-type s-wave exponential law \cite{p12}
\begin{equation}
\frac{\lambda_{ab}(T)-\lambda_{ab}(0)}{\lambda_{ab}(0)}=\sqrt{\frac{\pi\Delta_{0}}{2T}}exp(-\frac{\Delta_{0}}{k_{B}T}).
\end{equation}
but with a much smaller $\Delta_{0}$/k$_{B}$T$_{c}$ of 0.4 than
typical value of 1.76, indicating that isotropic s-wave model may
not be a good one for this new high-T$_{c}$ superconductor, although
a BCS-like gap is reported for SmFeAsO$_{0.85}$F$_{0.15}$ from
Andreev spectra \cite{p7}. However, the low temperature
$\lambda_{ab}$(T) can also be well fitted with a simple linear-T
dependence, indicating existence of nodal quasiparticle excitation,
possibly d-wave or extended s-wave pairing in nature \cite{p11}. A
triplet p-wave pairing model with strong ferromagnetic fluctuation
and degenerate electron Fermi surfaces was also proposed from band
structure consideration \cite{p10}. In the powder alignment
procedure, Fe spin-orbital-related anisotropic magnetic fluctuation
is needed at 300 K. However, we can not distinguish whether the
magnetic fluctuation is ferromagnetic or antiferromagnetic in
origin.

The anisotropic high-field ($\pm$7 T) isothermal superconducting
hysteresis loops M-B$_{a}$ at 5 K (Fig. 6) indicate that upper
critical field $\mu_{0}$H$_{c2}$ in both directions are much larger
than maximum applied field of 7 T with very short anisotropic
coherence lengths $\xi_{ab}$ and $\xi_{c}$. Moreover, hysteresis
loops with large paramagnetic background indicate the large
magnetic/impurity contribution.

\begin{figure}
\includegraphics{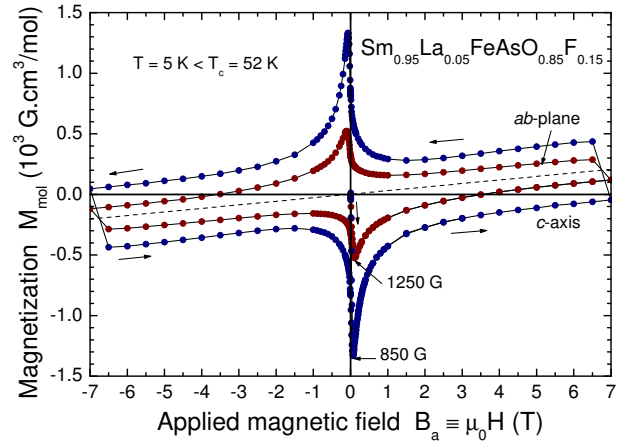}
\caption{\label{label}  Anisotropic high-field superconducting
hysteresis loop M$_{c}$(B$_{a}$) and M$_{ab}$(B$_{a}$) at 5 K for
aligned Sm$_{0.95}$La$_{0.05}$FeAsO$_{0.85}$F$_{0.15}$ powder.}
\end{figure}
In conclusion, due to Fe spin-orbital-related short-range
anisotropic exchange interaction $\chi_{ab}$(Fe) $> \chi_{c}$(Fe) at
300 K, Sm$_{0.95}$La$_{0.05}$FeAsO$_{0.85}$F$_{0.15}$
microcrystalline powder can be aligned using the field-rotation
alignment method where $\it{ab}$-plane is parallel to the aligned
field B$_{a}$ and $\it{c}$-axis is parallel to the rotation axis.
Rather small superconducting anisotropy parameter $\gamma \sim$ 2.4
+ 0.6 was observed (from both susceptibility and lower critical
field). The magnetic fluctuation used for powder alignment at 300 K
may be related with the pairing mechanism of superconductivity at
lower temperature.

This work was supported by NSC95-2112-M-007-056-MY3,
NSC97-2112-M-003-001-MY3, NSFC50421201, MSTC2006CB601003, and
MSTC2006CB922005.


\end{document}